\documentclass[aps,prl,floatfix,twocolumn,preprintnumbers,amsmath,amssymb,groupedaddress,showpacs,showkeys]{revtex4-1}

\usepackage{graphicx}  
\usepackage{dcolumn}   
\usepackage{bm}        
\usepackage{color}     
\usepackage{amsmath,amssymb}
\usepackage{psfrag}
\usepackage{bbold}
\usepackage{hyperref}

\usepackage{braket}  

\begin{document}
\title{Magnetoanisotropic Andreev Reflection in Ferromagnet/Superconductor Junctions} 
\author{Petra H\"{o}gl,$^1$ Alex Matos-Abiague,$^{1,2}$ Igor \v{Z}uti\'c,$^2$ and Jaroslav Fabian$^1$}
\affiliation{$^1$Institute for Theoretical Physics, University of Regensburg, 93040 Regensburg, Germany \\
	$^2$Department of Physics, University at Buffalo, State University of New York, Buffalo, NY 14260, USA}

\begin{abstract}
Andreev reflection spectroscopy of ferromagnet/superconductor (FS) junctions is an important probe of spin 
polarization. We theoretically investigate spin-polarized transport in FS junctions in 
the  presence of Rashba and Dresselhaus interfacial spin-orbit fields and show that Andreev reflection can be
controlled by changing the magnetization orientation. We predict a giant in- and out-of-plane magnetoanisotropy of
the junction conductance. If the ferromagnet is highly spin polarized---in the half-metal limit---the 
magnetoanisotropic Andreev reflection depends universally on the spin-orbit fields only. 
Our results show that Andreev reflection 
spectroscopy can be used for sensitive probing of interfacial spin-orbit fields in a FS junction.   
\end{abstract}

\maketitle

Spin-orbit coupling (SOC) is a key interaction in spintronics~\cite{Zutic2004:RMP,Fabian2007:APS,Maekawa:2006},
allowing an electrical control of magnetization and, vice versa, a magnetic control of electrical current. In systems
lacking space inversion symmetry---be it bulk, hybrid structures, junctions---SOC induces
spin-orbit fields~\cite{Zutic2004:RMP,Fabian2007:APS} as an emergent phenomenon. We are in particular concerned
here with interfacial spin-orbit fields which are believed to be behind a wealth of new phenomena, not 
existent or fragile in the bulk, such as the tunneling anisotropic magnetoresistance 
(TAMR)~\cite{Brey2004:APL, Gould2004:PRL,Moser2007:PRL,Matos-Abiague2009:PRB,*Matos-Abiague2009:PRB2}, 
interfacial spin-orbit torques~\cite{Ryu2013:NN,*Chernyshov2009:NP}, or Skyrmions~\cite{Fert2013:NN}. 

Interfacial spin-orbit fields are also important in semiconductor/superconductor~\cite{Lutchyn2010:PRL, Oreg2010:PRL, Mourik2012:S, Rokhinson2012:NP} 
and ferromagnet/superconductor (FS) 
junctions~\cite{Nadj-Perge2014:S} for creating Majorana quasiparticle states. It is the latter junctions that we focus on. We investigate
the interplay of magnetism and spin-orbit fields. We show that this interplay leads to marked anisotropies in the junction 
conductance  with respect to the orientation of magnetization. The most robust is the out-of-plane anisotropy (plane being the interface),
which arises from the omnipresent Rashba field~\cite{Bychkov:1984}. A more subtle is the in-plane anisotropy, which arises from the
interference between the Rashba and Dresselhaus~\cite{Dresselhaus:1955} fields, induced by a twofold anisotropy of the $C_{2v}$ type.
A zinc-blende semiconductor (say, GaAs or InAs) as a barrier in an FS junction would create such an anisotropy, generating spin-orbit fields $C_{2v}$ ``butterfly'' patterns, as shown by first-principles calculations~\cite{Gmitra2013:PRL}.   
Remarkably, the resulting magnetoconductance anisotropy---we term it {\it magnetoanisotropic 
Andreev reflection (MAAR)}---is giant in comparison to TAMR, its
normal-state counterpart, reaching a universal behavior in the half-metallic case. This is because  Andreev reflection (AR) 
(which has no counterpart in the normal-state TAMR) is strongly influenced by interfacial spin-orbit fields.

We specifically examine the influence of SOC and crystalline anisotropy on the process of AR in which the reflected particle carries the information about both the
phase of the incident particle and the macroscopic phase of the superconductor to which
a Cooper pair is being transferred~\cite{Deutscher2005:RMP}. AR is thus responsible for 
the proximity effect in which the phase correlations are introduced to a nonsuperconducting 
material~\cite{Buzdin2005:RMP,Bergeret2005:RMP,Golubov2004:RMP,Eschrig2011:PT,Visani2012:NP}. 
While the main interest in AR is currently the proximity effect coupled with SOC, inducing Majorana
states, in spintronics AR is used to probe the spin polarization in FS
junctions~\cite{Deutscher2005:RMP,Buzdin2005:RMP,Bergeret2005:RMP,Golubov2004:RMP,%
Eschrig2011:PT,Visani2012:NP,Kashiway2000:RPP,deJong1995:PRL,Soulen1998:S,Upadhyay1998:PRL,%
Zhu1999:PRB,Kashiwaya1999:PRB,Zutic1999:PRB,Zutic1999:PRB2,*Zutic2000:PRB,%
Mazin2001:JAP,Kikuchi2001:PRB,Turel2011:APL}. We argue that AR can also be a sensitive probe of 
interfacial spin-orbit fields.

\begin{figure}[htp]
{\includegraphics[width=0.8\columnwidth]{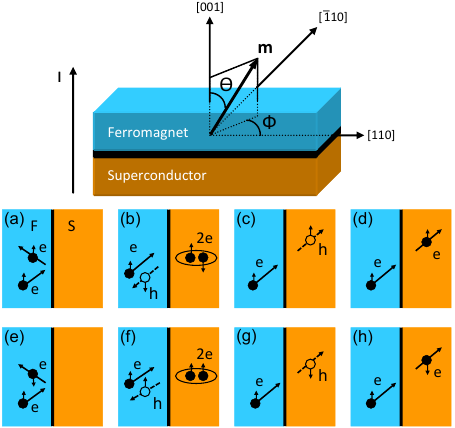}}
	\caption{(Color online) Top: FS junction.  Magnetization vector $\boldsymbol{m}$ is given by the 
polar angle $\Theta$ and azimuthal angle $\Phi$. Current, $I$, flows  perpendicular to the 
interface. To specify spin-orbit fields we use principal crystallographic 
orientations $x=[110]$, $y=[\overline{1}10]$, and $z=[001]$.
Bottom: Scattering processes at the FS interface with SOC. Electrons (holes) are depicted by full   
		(empty) circles.  Vertical arrows denote the spin. The processes for a spin up incoming electron:  (a) Specular reflection, (b) Andreev reflection, (c) holelike transmission, 
and (d) electronlike transmission.  (e)-(h) Corresponding spin-flip counterparts.  
Spin-flip (equal electron and hole spins) Andreev reflection is in (f).}
	\label{fig:fmsc}
\end{figure}

Our model FS junction consists of F ($z<0)$ and S ($z>0$) semi-infinite regions separated by a flat interface at $z=0$,
with potential and SOC scattering. The scheme and possible scattering channels are illustrated in Fig.~\ref{fig:fmsc}. 
For example, in conventional AR the incoming electron is reflected as a hole with the opposite spin, while 
spin-flip AR implies equal spin of the incoming and reflected particles. These two AR processes, see 
Figs.~\ref{fig:fmsc}(b) and~\ref{fig:fmsc}(f), introduce, respectively, spin-singlet and spin-triplet superconducting correlations at the 
interface~\cite{Eschrig2011:PT,Visani2012:NP}.

We consider epitaxial-quality junctions, such as those used in TAMR~\cite{Moser2007:PRL}, or point contact geometries~\cite{Stamenov2013:JAP,*Stamenov2013:JAP2}, in which ballistic transport formalism is applicable. In diffusive tunnel junctions AR could be enhanced by electron-hole coherence~\cite{Hekking1993:PRL}. In ferromagnetic junctions such effects would be absent for normal AR due to short coherence length, but 
spin-flip AR could be enhanced. (Ordinary effects of diffusion could be accounted for by 
renormalizing the tunneling parameters~\cite{Mazin2001:JAP}).
We generalize the Blonder-Tinkham-Klapwijk formalism~\cite{Blonder1982:PRB} and solve the Bogoliubov-de~Gennes equation~\cite{deGennes:1989} for 
quasiparticle states $\Psi(\mathbf{r})$ with energy $E$,
\begin{eqnarray}
	\begin{pmatrix} \hat{H}_e & \hat{\Delta} \\ \hat{\Delta}^\dagger & \hat{H}_h  \end{pmatrix}\Psi(\mathbf{r})
	&=& E \Psi(\mathbf{r}),
	\label{eq:2}
\end{eqnarray}
with the single-particle Hamiltonian for electrons $\hat{H}_e=-(\hbar^2/2) \boldsymbol{\nabla}\left[1/m(z)\right]\boldsymbol{\nabla}-\mu(z)- (\Delta_{xc}/2) \Theta(-z) \mathbf{m}\cdot\boldsymbol{\hat{\sigma}}+ \hat{H}_{B}$;  
for holes $\hat{H}_h=-\hat{\sigma}_y\hat{H}_e^*\hat{\sigma}_y$. The unit magnetization vector (see 
Fig.~\ref{fig:fmsc}) is
$\mathbf{m}=\left[\sin\Theta \cos\Phi, \sin\Theta \sin\Phi, \cos\Theta\right]$, 
$\boldsymbol{\hat{\sigma}}$ are Pauli matrices, 
$\Delta_{xc}$ is the exchange spin splitting in the F region (Stoner model), 
$m(z)$ is the effective mass, and $\mu(z)$ is the chemical potential. 
The interfacial scattering 
is modeled as $\hat{H}_B=\left(V_0 d+\mathbf w \cdot \boldsymbol{\hat{\sigma}}\right) \delta(z)$,
where $V_0$ and $d$ are the barrier height and width, while
 $\mathbf w =[(\alpha-\beta) k_{y}, -(\alpha+\beta) k_{x},0]$ 
is the effective SOC field including Rashba and 
Dresselhaus terms~\cite{Zutic2004:RMP,Fabian2007:APS}, parametrized by 
$\alpha$ and $\beta$, respectively, for the crystallographic orientations see Fig.~\ref{fig:fmsc}.
The superconducting pair potential is given by 
$\hat{\Delta}= \Delta\Theta(z) \mathbb{1}_{2\times2}$ (the accuracy of such a step-function form 
of $\hat\Delta $ is discussed  in Ref.~\cite{Beenakker1997:RMP}), with the isotropic gap $\Delta$.
Similar methodology, for half-metal/S junctions with Rashba coupling inside the superconductor
was employed in Ref.~\cite{Duckheim2011:PRB}. With Rashba-only SOC one should still
obtain out-of-plane magnetoanisotropy, and this is already implicit in this formalism 
\cite{Duckheim2011:PRB}.

Since the in-plane wave vector $\mathbf{k}_{||}$ is conserved, 
$\Psi_{\sigma}(\mathbf{r})=\Psi_{\sigma}(z) e^{i\mathbf{k_{||}}\mathbf{r_{||}}}$. The solution in the F region for 
incoming electrons with spin $\sigma$ is
\begin{eqnarray}
\Psi^{F}_{\sigma}&=&\frac{1}{\sqrt{k^e_{\sigma}}} e^{i k^e_{\sigma} z} \chi^e_{\sigma}+r^e_{\sigma, \sigma} e^{-i k^e_{\sigma} z} \chi^e_{\sigma}+ r^e_{\sigma, -\sigma} e^{-i k^e_{-\sigma} z} \chi^e_{-\sigma}  \nonumber \\
& & + r^h_{\sigma, -\sigma} e^{i k^h_{-\sigma} z} \chi^h_{-\sigma} +r^h_{\sigma, \sigma} e^{i k^h_{\sigma} z} \chi^h_{\sigma},
\label{eq:8}
\end{eqnarray}
with the spinors for the electronlike $\chi^e_{\sigma}=\left(\chi_{\sigma},0\right)^T$ and holelike $\chi^h_{\sigma}=\left(0,\chi_{-\sigma}\right)^T$ quasiparticles, 
both containing
\begin{eqnarray}
\chi_\sigma^T=
\left(\sigma\sqrt{1+\sigma \cos\Theta} e^{-i\Phi},
\sqrt{1-\sigma \cos\Theta}\right)/\sqrt{2},
\label{eq:10}
\end{eqnarray}
where $\sigma=1(-1)$ corresponds to the spin parallel (antiparallel) to $\mathbf{\hat{m}}$. 
The electronlike (holelike) quasiparticle wave vectors in the F region are $k^{e(h)}_{\sigma}=\sqrt{k_F^2+2m_{F}/\hbar^2\left[(-)E+ \sigma\Delta_{xc}/2\right]-k^2_{||}}$. 

In the S region the scattering states are
\begin{eqnarray} \nonumber
\Psi^{S}_{\sigma}&=& t^e_{\sigma, \sigma} e^{i q^e z}
\begin{pmatrix}
u \\
0\\
v \\
0
\end{pmatrix}
+t^e_{\sigma, -\sigma} e^{i q^e z}
\begin{pmatrix}
0 \\
u \\
0 \\
v
\end{pmatrix}\\
&+&  t^h_{\sigma, \sigma} e^{-i q^h z}
\begin{pmatrix}
v \\
0 \\
u  \\
0
\end{pmatrix}
+  t^h_{\sigma, -\sigma} e^{-i q^h z} 
\begin{pmatrix}
0 \\
v \\
0 \\
u 
\end{pmatrix},
\label{eq:14}
\end{eqnarray}
where the quasiparticle wave vectors are given by $\nolinebreak{q^{e(h)}=\sqrt{q_F^2+(-)2m_{S}/\hbar^2\sqrt{E^2-\Delta^2}-k^2_{||}}}$. 
The superconducting coherence factors satisfy  $u^2=1-v^2= \left(1+ \sqrt{E^2-\Delta^2}/E\right)/2$.

Using charge current  conservation,  the differential conductance at zero temperature, normalized by the Sharvin 
conductance~\cite{Zutic2004:RMP} $G_{Sh}=e^2k_F^2A/ (2\pi h)$ of a perfect contact, is
\begin{eqnarray}
G=  \sum_{\sigma} \int{ \frac{d^2 \mathbf k_{\|}}{2\pi k_F^2}  \left[1+R^h_\sigma(-e\mathrm{V})-R^e_\sigma(e\mathrm{V})\right]},
\label{eq:21}
\end{eqnarray}
containing the probability amplitudes in the F region
$\nolinebreak{R^{e(h)}_{\sigma}(E,\mathbf{k_{\|}})=
\mathrm{Re}\left( k^{e(h)}_{\sigma} 
\left|r^{e(h)}_{\sigma,\sigma}\right|^2+k^{e(h)}_{-\sigma} \left|r^{e(h)}_{\sigma,-\sigma}\right|^2 \right)}$
which combine the coefficients for 
the scattering processes with and without spin flip for specular reflection and AR; $V$ is the bias voltage and $A$ is the  interfacial area.
\begin{figure}[h] 
 \includegraphics[width=0.8\columnwidth]{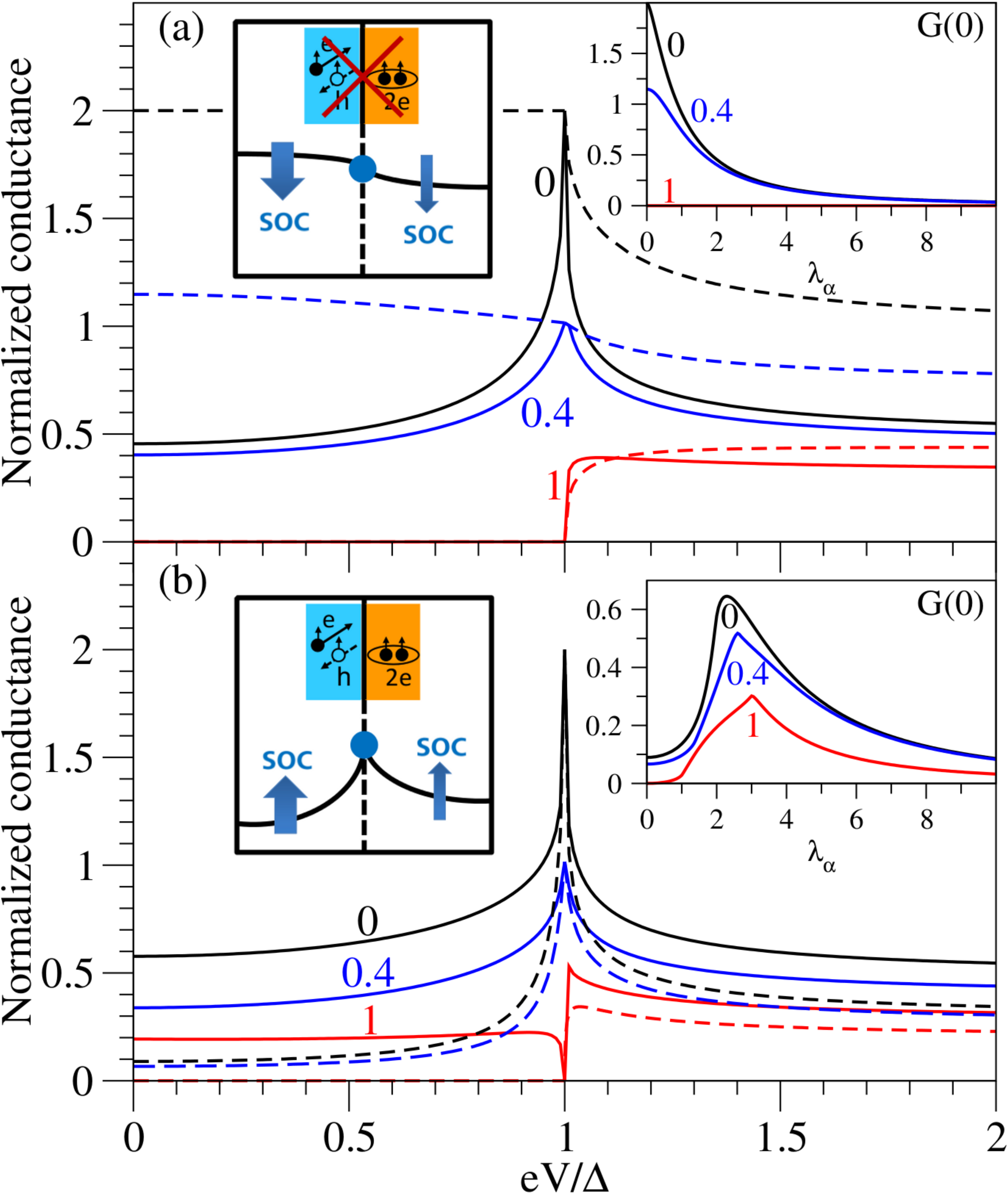}%
 \caption{\label{fig:deltaxc}(Color online) Calculated normalized conductance, $G(eV/\Delta)$, for
different (indicated) spin polarizations $P$. Rashba SOC is $\lambda_\alpha=2$ 
and Dresselhaus SOC is $\lambda_\beta=0$. Magnetization is in plane ($\Theta=\pi/2$). (a) No interfacial barrier  
($Z=0$), and  (b) modest interfacial barrier ($Z=1$) cases are shown. 
The dashed lines show $G$ without SOC. The insets show the dependence of  $G(0)$ on Rashba SOC. 
In (a),  the subgap conductance for $P=1$ vanishes for every $\lambda_{\alpha,\beta}$; 
in the inset $G(0)=0$ for this case. The illustrations summarize main qualitative impacts of SOC on conductance.} 
\end{figure}

To describe our results we introduce dimensionless quantities:
$Z= V_0d\sqrt{m_{F}m_{S}} /\left(\hbar^2 \sqrt{k_Fq_F}\right)$ denotes the 
barrier strength~\cite{Blonder1982:PRB,Zutic1999:PRB2}, 
$\lambda_\alpha=2\alpha\sqrt{m_{F}m_{S}}/\hbar^2$ and $\lambda_\beta=2\beta\sqrt{m_{F}m_{S}}/\hbar^2$
quantify the Rashba and Dresselhaus SOC, and $P=\left(\Delta_{xc}/2\right)/\mu_{F}$ defines
the spin polarization in F.

We first examine the influence of SOC on the FS conductance (see Fig.~\ref{fig:deltaxc}), for
a metallic point contact ($Z=0$) and for a moderate barrier ($Z=1$). For the former case the conductance
tends to decrease with increasing SOC. Even in the half-metallic case ($P=1$) SOC does not
give a finite subgap conductance; spin-flip AR is suppressed. In contrast, for moderate
barrier ($Z=1$),  SOC enhances the conductance due to spin-flip AR,  even for $P=1$.
Interestingly, at $\mathrm{eV}= \Delta$ the conductance is not affected by SOC for any $Z$. 
Focusing on $G(0)$, Fig.~\ref{fig:deltaxc} shows that in a metallic 
contact increasing SOC steadily reduces $G(0)$, while for a moderate barrier 
 $G(0)$ is a nonmonotonic  function of SOC, with a ($P$-dependent) maximum which 
turns out to be due to spin-flip AR.

The absence of spin-flip AR in metallic contacts can be explained
analytically. For $\mathrm{eV}\le\Delta$ quasiparticle transmission is prohibited 
and subgap conductance 
$G \sim \sum_{\sigma} \int d^2 \mathbf k_{\|} \,2R^h_\sigma(-e\mathrm{V})$. 
In the half-metallic case the only contribution to AR comes from spin-flip 
AR, $R^h_{\sigma}(E)=\mathrm{Re}\left( k_{\sigma}^h \left|r^h_{\sigma,\sigma}\right|^2 \right)$,
because of the missing minority spin subband in F. To lowest order in SOC and $Z$
then  $G(0)\propto Z^2\lambda_i^2$ with $i\in \{\alpha,\beta\}$~\cite{Supplementary}, vanishing
if $Z=0$. This perturbative quadratic dependence on the spin-orbit strength was also obtained in 
Ref.~\cite{Duckheim2011:PRB}.
    
The calculated conductance features of SOC~\cite{Gorkov2001:PRL,Mizuno2009:PRB,Linder2011:PRL,Dolcini2008:PRB,Sun2015:PRB} 
can be distinguished from $k$-{\it independent} spin-flip scattering by magnetic moments: 
For $Z=0$ SOC always reduces the conductance and the subgap conductance vanishes
for $P=1$. In contrast, $k$-independent spin-flip scattering ~\cite{Balatsky2006:RMP} 
can increase the conductance and the subgap conductance is in general finite for $P=1$. However, similar features
as those of SOC can arise in exotic superconductors without bulk inversion 
symmetry~\cite{Wu2009:PRB,*Wu2010:PRB}.

\begin{figure}[hb]
\includegraphics[width=0.75\columnwidth]{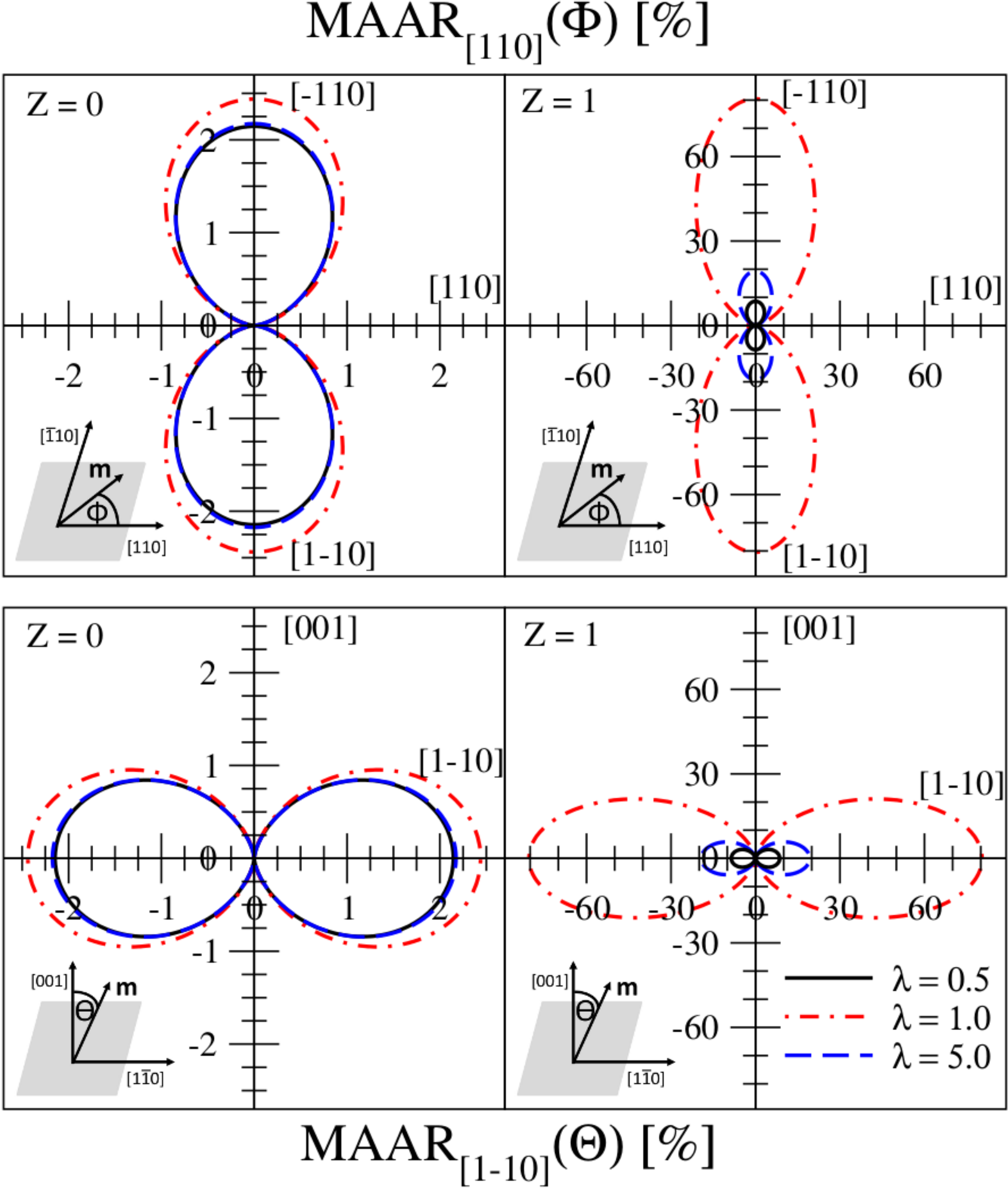}%
\caption{\label{fig:AAR}(Color online) Top:  
Calculated in-plane magnetoanisotropic Andreev reflection (MAAR)   
with $[110]$ crystallographic reference axis for $Z=0$ (left) and $Z=1$ (right)
 for different strengths of SOC $\lambda_\alpha=\lambda_\beta=\lambda$ at $P=0.4$ and $V=0$.
Bottom: Out-of-plane MAAR with $[1\overline{1}0]$ crystallographic reference axis.
For $Z=0$ the lines of $\lambda=0.5$ and $\lambda=5.0$ coincide. For the chosen reference axes 
and $\lambda_\alpha=\lambda_\beta$ the in-plane and out-of-plane MAAR curves have the 
same magnitude and shape, but rotated to the corresponding reference axis.}
 \end{figure}

While the conductance changes are indicative of interfacial SOC, magnetic
anisotropy of the conductance is a true fingerprint. As the main contribution comes
from AR, we call this anisotropy effect {\it magnetoanisotropic Andreev reflection}. 
We consider two configurations: in plane, in which magnetization $\mathbf{m}$ changes
azimuthally ($\Phi$) in the interfacial plane, and out of plane, with polar ($\Theta$) changes of  $\mathbf{m}$
in a perpendicular plane (see Fig. 1). 
We define the in-plane MAAR as
\begin{eqnarray}
	\mathrm{MAAR}_{[110]}(\Phi)&=& \left.\frac{G(\Theta,0)-G(\Theta,\Phi)}{G(\Theta,\Phi)}\right|_{\Theta=90^\circ},
	\label{eq:ip}
\end{eqnarray}
and the out-of-plane MAAR as
\begin{eqnarray}
	\mathrm{MAAR}_{[1\overline{1}0]}(\Theta)&=& \left.\frac{G(0,\Phi)-G(\Theta,\Phi)}{G(\Theta,\Phi)}\right|_{\Phi=-90^\circ}.
	\label{eq:oop}
\end{eqnarray}
The out-of-plane MAAR depends, in general on $\Phi$, but we choose the $yz$ ($\Phi=-90^\circ$) 
plane as its reference.

The calculated MAAR, in Fig.~\ref{fig:AAR}, shows a nonmonotonic dependence on SOC. 
For metallic contacts ($Z=0$) MAAR is determined by the magnetoanisotropy of conventional AR. 
In the presence of a barrier (exemplified by $Z=1$),  MAAR gets strongly enhanced due to 
the additional contribution from spin-flip AR. In-plane MAAR exhibits 
$C_{2v}$ symmetry due to the interplay of Rashba and Dresselhaus fields, 
similarly to TAMR~\cite{Matos-Abiague2009:PRB,Matos-Abiague2009:PRB2,Fabian2007:APS,Moser2007:PRL}.
If either of the two fields is absent, in-plane MAAR vanishes. In contrast, out-of-plane MAAR
is finite even with the Rashba field alone, which makes it a robust probe of this important interfacial SOC. 
Interestingly, at $\mathrm{eV}=\Delta$ MAAR is always 
absent, as there are no effects of SOC on $G$ here; see the discussion to Fig. 2. 
Additional effects (such as appearance
of symmetry lobes) can arise due to the effective mass and Fermi wave vector mismatch ~\cite{Supplementary}.

\begin{figure}[b]
	\includegraphics[width=0.8\columnwidth]{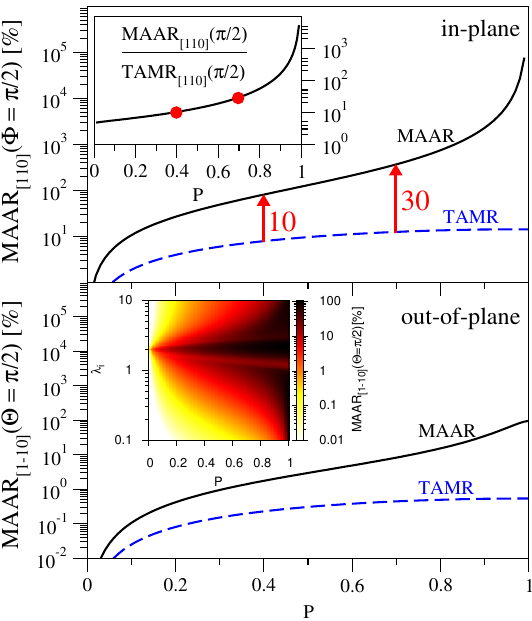}%
	\caption{\label{fig:map} (Color online) Calculated in-plane (top) and out-of-plane (bottom)  MAAR and TAMR as a function of spin polarization $P$ for
a moderate barrier ($Z=1$) and $V=0$. The in-plane case is calculated with $\lambda_\alpha = \lambda_\beta=1$, while for out of plane we
have included Rashba $\lambda_\alpha=0.5$  only. The top inset shows
the ratio of MAAR and TAMR for the in-plane case, while the bottom inset shows the color map of out-of-plane MAAR as a function of $P$
and Rashba (or Dresselhaus) SOC $\lambda_i$ (where $i$
could be either $\alpha$ or $\beta$). }
\end{figure}

{\it Compared to TAMR, the magnitude of MAAR is giant}, varying by orders of magnitude upon
changing the spin polarization $P$. (The experimentally measured in-plane TAMR in Fe/GaAs/Au junctions 
is less than a percent~\cite{Moser2007:PRL}.) A detailed model comparison is shown in Fig. 4 for
both in- and out-of-plane configurations; 
TAMR is evaluated by setting $\Delta = 0$. For a typical $P$ of 40\%, the ratio MAAR/TAMR is about 10. Moving towards half metals ($P \agt 80\%$), this ratio climbs to more than $10^2$. 
This giant increase is best illustrated in the half-metallic limit of $P=1$. For a weak SOC (which is
typically the case)
an analytical treatment gives~\cite{Supplementary},  
\begin{eqnarray}
	\mathrm{MAAR}_{[110]}(\!\Phi\!)&\!\approx\!&\frac{2\lambda_\alpha\lambda_\beta(1-\cos2\Phi)}{\lambda_\alpha^2+\lambda_\beta^2+2\lambda_\alpha\lambda_\beta\cos2\Phi},
\end{eqnarray}
\begin{eqnarray}
	\hspace*{-0.5cm}\mathrm{MAAR}_{[1\overline{1}0]}(\!\Theta\!)&\!\approx\!&\frac{(\lambda_\alpha+\lambda_\beta)^2(1-\cos2\Theta)}{3(\!\lambda_\alpha^2\!+\!\lambda_\beta^2\!)\!-\!2\lambda_\alpha\lambda_\beta\!+\!(\!\lambda_\alpha\!+\!\lambda_\beta\!)^2\!\cos2\Theta}.
	\label{eq:oop_app}
\end{eqnarray}
Therefore, the in-plane $\mathrm{MAAR}_{[110]}(\Phi=\pi/2)\approx 4\lambda_\alpha \lambda_\beta/(\lambda_\alpha-\lambda_\beta)^2$, and
out-of-plane $\mathrm{MAAR}_{[1\overline{1}0]}(\Theta=\pi/2)\approx (\lambda_\alpha+ \lambda_\beta)^2/(\lambda_\alpha-\lambda_\beta)^2$, {\it depending universally on 
the spin-orbit fields only},  and {\it diverging} as 
$\lambda_\alpha \approx \lambda_\beta$ (see the in-plane
case in Fig. 4).  In contrast, TAMR, which is proportional  to the product $\lambda_\alpha \lambda_\beta$~\cite{Matos-Abiague2009:PRB}, has no singular behavior, and is not 
a universal function of $\lambda_i$ only. 

We can trace this giant enhancement of MAAR over TAMR to spin-flip AR. 
Let us separate phenomenologically the conductance $G=G^{(0)}+G_{so}$ into the sum of SOC independent and dependent parts. 
In TAMR typically $G^{(0)}\gg G_{so}$, and  TAMR $\sim G_{so}/G^{(0)} \ll 1$, even for $P\approx 1$. 
But in FS junctions $G^{(0)}$ decreases with increasing $P$, eventually vanishing in the half-metallic limit. 
For $P \approx 1$ the conductance of the FS junction is dominated by the spin-flip AR contribution to $G_{so}$. 
Thus, SOC determines both the conductivity and the magnetoanisotropy. Furthermore, if 
$\lambda_\alpha\approx \pm\lambda_\beta$, the spin-flip AR, and so the conductance, can be switched {\em on} 
and {\em off} by changing the orientation of  $ \mathbf {m}$.  
For $\lambda_\alpha =\lambda_\beta$ and $\Phi = 0$, $\mathbf{m}\perp \mathbf{w}$ and spin-flip AR yields a finite $G$. 
However, if $\Phi = \pi/2$, then $\mathbf{m}\parallel\mathbf{w}$ and spin-flip processes are strongly suppressed;
$G(eV \leq \Delta) $ at $\Phi = \pi/2$ vanishes.  As a result, {\it in-plane MAAR diverges if $\lambda_\alpha =\lambda_\beta$}. 
Similarly for out-of-plane MAAR.

There is one more peculiarity of MAAR in the half-metallic limit. If only Rashba (or only Dresselhaus) SOC is present, out-of-plane MAAR has a {\it fixed universal} magnitude of 100\%.  This is shown 
in Fig. 4 (in particular the inset for $\lambda_i \alt 1$ shows 
MAAR of 100\% for $P\approx 1$).
It follows from  Eq.~(\ref{eq:oop_app}) that 
$\mathrm{MAAR}_{[1\overline{1}0]}(\Theta)\approx (1-\cos 2\Theta)/(3+\cos 2\Theta)$, 
which gives a universal amplitude of $100\%$ for $\Theta=\pi/2$. In other words, 
$G(\Theta=0)=2G(\Theta=\pi/2)$. The origin of this universal behavior is traced to the 
spin-flip probability by scattering of spin-polarized electrons off spin-orbit fields.  
The conductance is determined by spin-flip AR. For out-of-plane magnetization, $\Theta=0$,
two fields, one along $x$ and one along $y$, induce a spin flip. But for an in-plane
magnetization, say along $x$,  $\Theta=\pi/2$, only the spin-orbit field component 
along $y$ can flip  the spin. This gives the $2:1$ ratio in conductances and $100\%$ of MAAR.
 A more technical and detailed discussion of
the differences between MAAR and TAMR can be found in Ref.~\onlinecite{Supplementary}.

Experimental realization of MAAR could follow the measurement geometry of TAMR~\cite{Moser2007:PRL},
ideally also the same junction, with the nonmagnetic metal that becomes superconducting 
at low temperatures. Magnetization of the ferromagnetic layer is typically rotated by an 
external magnetic field. This field can  bring additional anisotropic orbital effects whose presence 
can be clearly identified from the field magnitude dependence~\cite{Wimmer2009:PRB}. However, one can avoid 
these extrinsic effects entirely if one uses dysprosium magnets which can be oriented 
by the field, but do not need its presence to remain in the rotated position~\cite{Betthausen2012:Science}. 
Potential aspects of nonflat tunneling barriers can also be treated ~\cite{Chen2001:PRB}.
A practical alternative (especially if ballistic junctions are desired) could be a point 
contact FS junction geometry~\cite{Stamenov2013:JAP,Stamenov2013:JAP2,Alicea2012:RPP}.

To conclude, we have applied a well-established theoretical formalism to systematically explore
the magnetic anisotropy of the conductance in FS junctions due to interfacial SOC. 
We predict a giant in- and out-of-plane MAAR---when compared with TAMR---exhibiting universal 
characteristics in  the half-metallic regime. The predicted magnetization control of the 
AR suggests a similar control of the superconducting proximity effect and 
Majorana states. Our findings reveal an unexplored venue 
for AR spectroscopy, in the sensitive probing of interfacial SOC and 
related magnetoanisotropic phenomena.

We acknowledge useful discussions with D. Weiss, C. Strunk, C. Back, B. Nadgorny, and P. Stamenov.
This work has been supported by the DFG SFB 689, International Doctorate Program
Topological Insulators of the Elite Network of Bavaria,
DOE-BES Grant No. DE-SC0004890 (I.\v{Z}.),  and ONR N000141310754 (A.M.). 

\bibliography{references}

\newpage

\begin{figure}[t]
	\includegraphics[width=\textwidth]{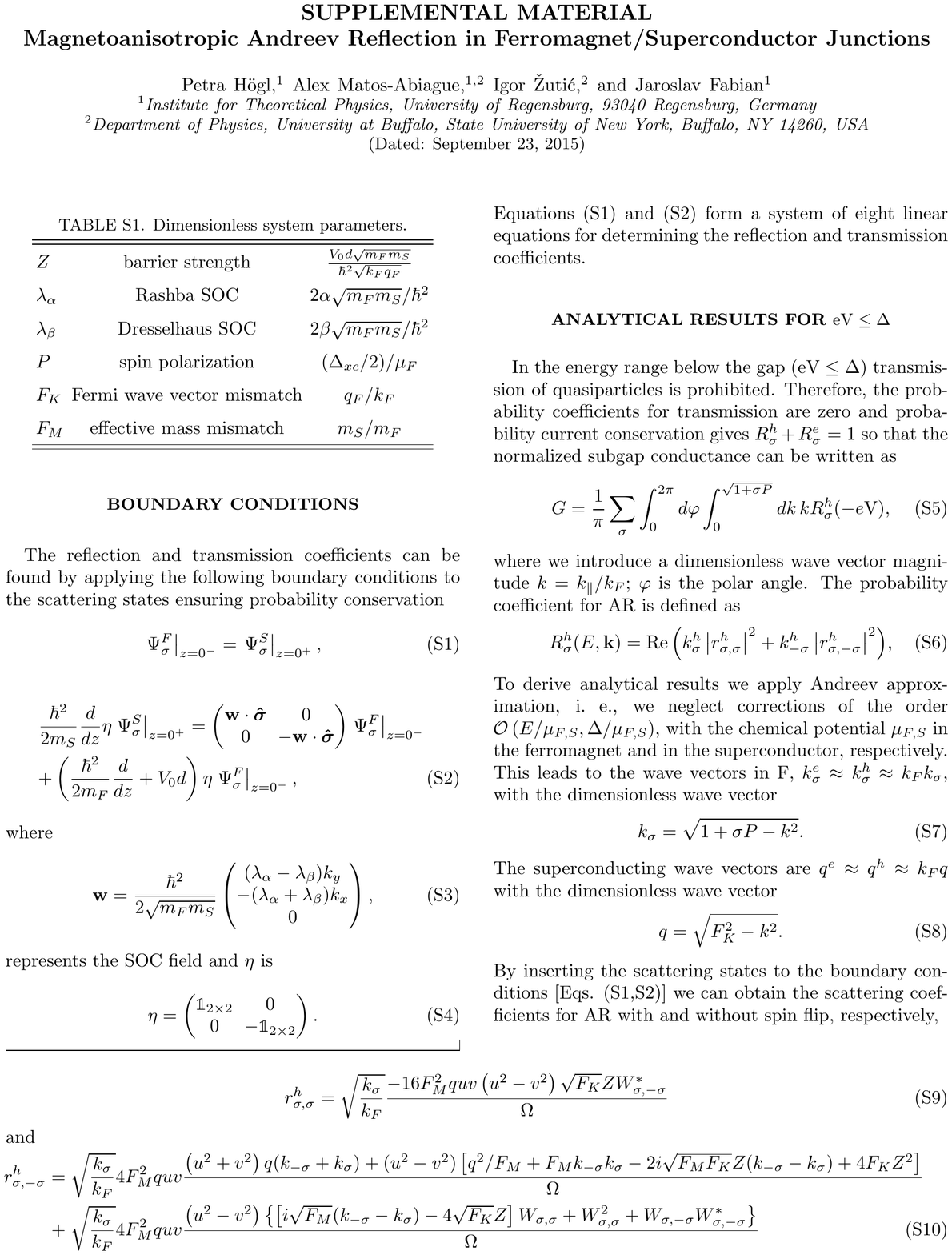}%
\end{figure}
\begin{figure}[t]
	\includegraphics[width=\textwidth]{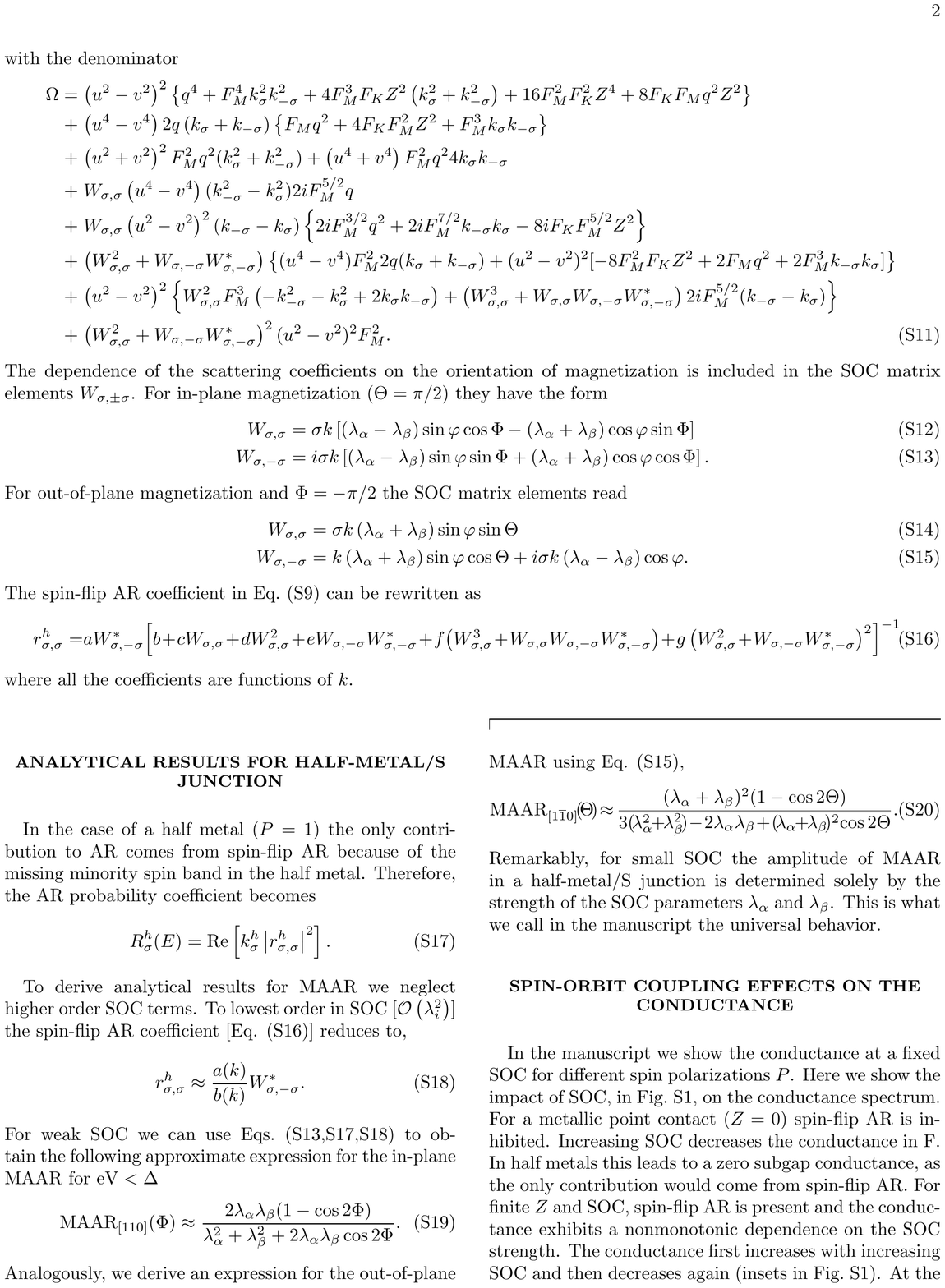}%
\end{figure}
\begin{figure}[t]
	\includegraphics[width=\textwidth]{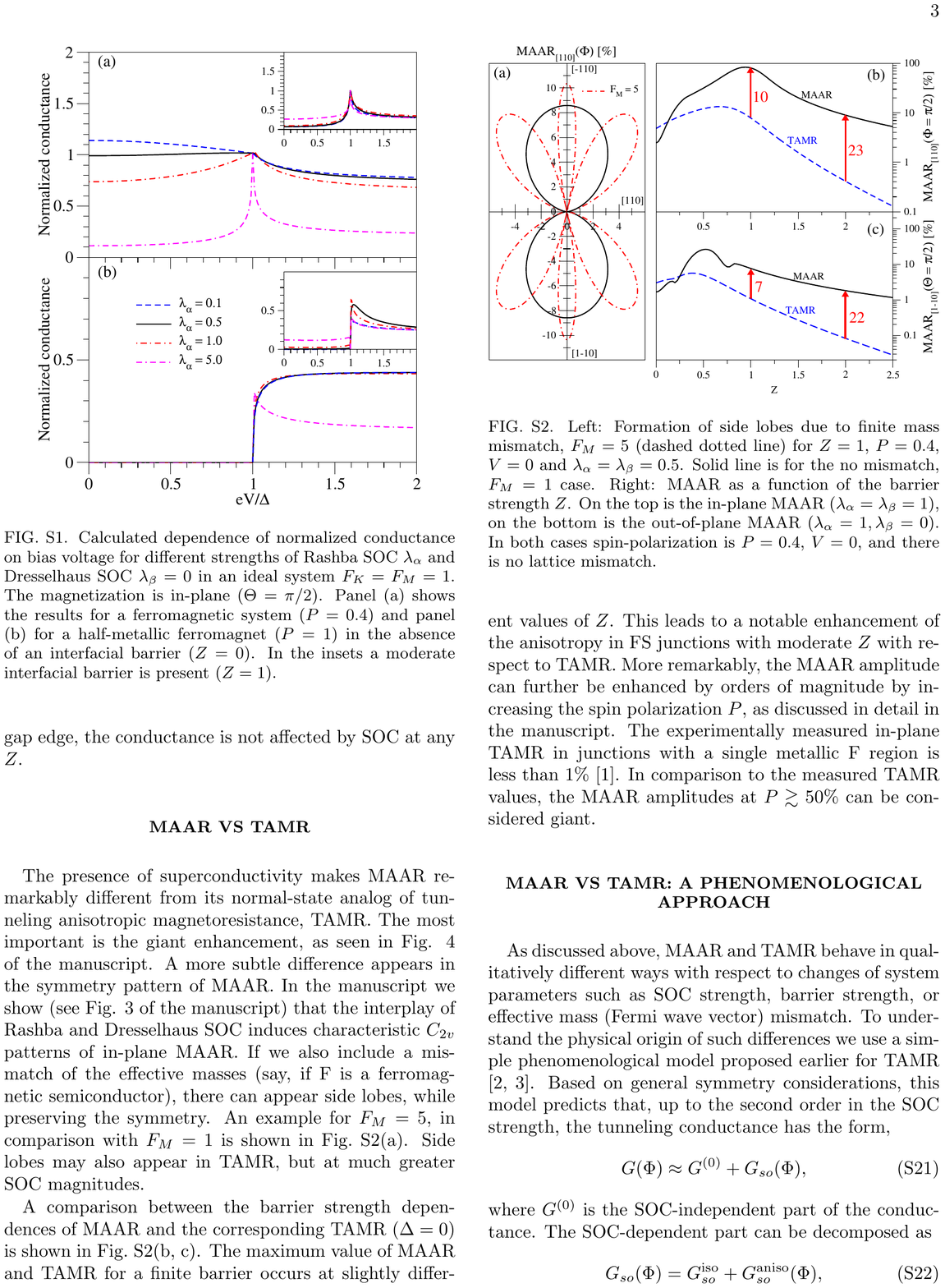}%
\end{figure}
\begin{figure}[t]
	\includegraphics[width=\textwidth]{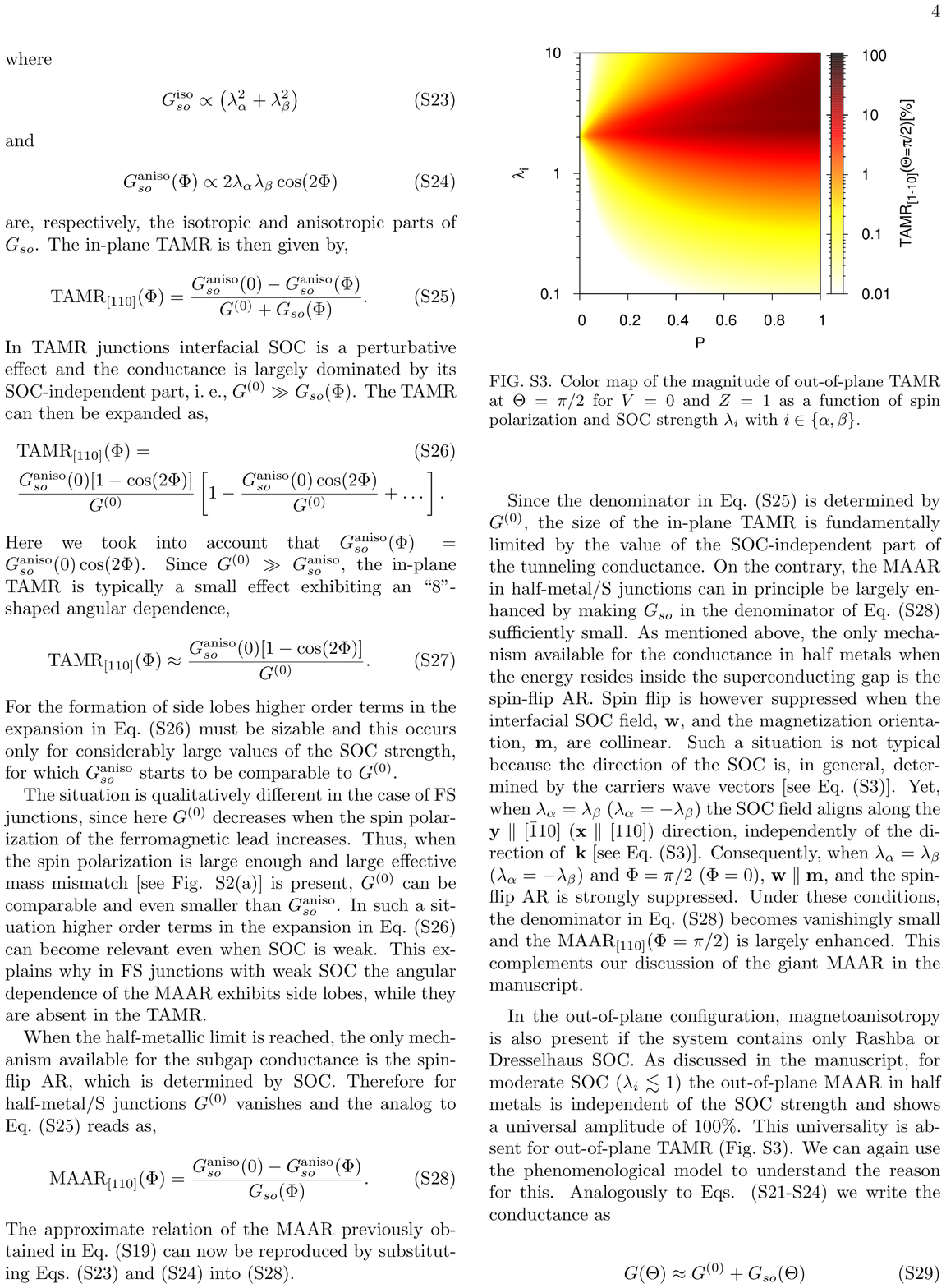}%
\end{figure}
\begin{figure}[t]
	\includegraphics[width=\textwidth]{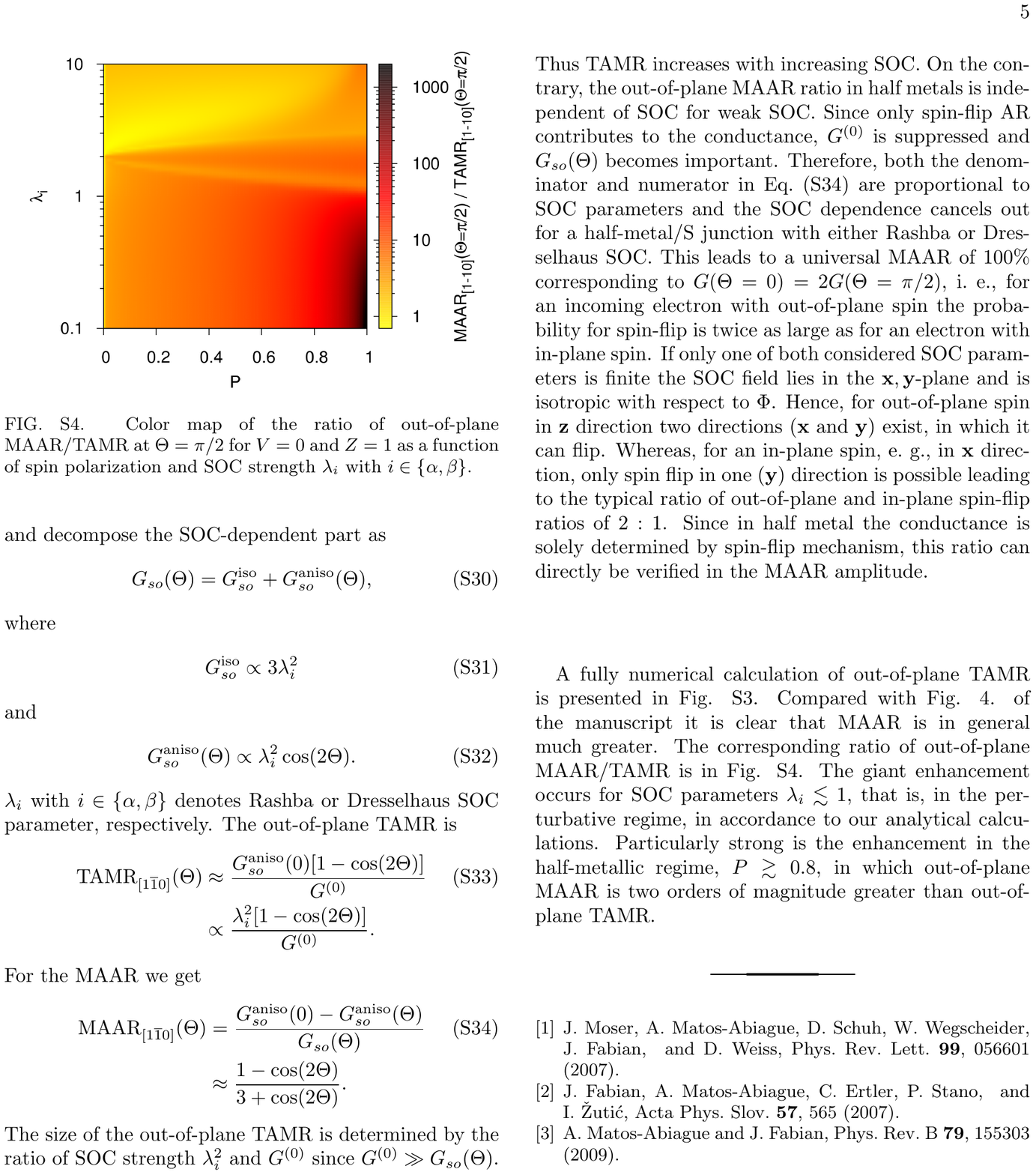}%
\end{figure}

\pagestyle{empty}

\end{document}